\documentclass[iop]{emulateapj}
\usepackage{graphicx}
\usepackage{natbib}
\usepackage{amsmath}
\usepackage{color}
\usepackage{ulem}
\usepackage[caption=false]{subfig} % mac
\usepackage{subfloat}
\slugcomment{{\sc Published in ApJ, Vol. 851:25 (2017)}}
\newcommand{\added}[1]{#1}

\makeatletter
\newcommand{\Yd}{X_{\rm D}}
\newcommand{\Ydeq}{X_{\rm D,eq}}
\newcommand{\Ydp}{X_{\rm D}^{\rm P}}
\newcommand{\dhr}{({\rm D}/{\rm H})}
\newcommand{\dhrp}{({\rm D}/{\rm H})_{\rm P}}
\newcommand{\dhrism}{({\rm D}/{\rm H})_{\rm ISM}}
\newcommand{\Md}{M_{\rm D}}
\newcommand{\Mddot}{\dot{M}_{\rm D}}
\newcommand{\hethree}{^3{\rm He}}
\newcommand{\hehp}{(^3{\rm He}/{\rm H})_{\rm P}}
\newcommand{\frec}{f_{\rm rec}}
\newcommand{\fprim}{f_{\rm prim}}

\newcommand{\mocc}{m_{\rm O}^{\rm cc}}

\newcommand{\Mo}{M_{\rm O}}

\newcommand{\Modot}{\dot{M}_{\rm O}}

\newcommand{\mdotstar}{\dot{M}_*}
\newcommand{\mdotstaro}{\dot{M}_{*,0}}
\newcommand{\mdotstart}{\dot{M}_*(t)}

\newcommand{\mdotout}{\dot{M}_{\rm out}}
\newcommand{\mdotinf}{\dot{M}_{\rm inf}}

\newcommand{\Yo}{Z_{\rm O}}

\newcommand{\Yoeq}{Z_{\rm O,eq}}

\newcommand{\ohh}{[{\rm O}/{\rm H}]}

\newcommand{\taustar}{\tau_*}

\newcommand{\tausfh}{\tau_{\rm sfh}}
\newcommand{\taudep}{\tau_{\rm dep}}

\newcommand{\taubar}{\bar{\tau}}

\newcommand{\taubarDepSfh}{\bar{\tau}_{[{\rm dep,sfh}]}}

\newcommand{\Gyr}{\,{\rm Gyr}}

\newcommand{\yr}{\,{\rm yr}}

\makeatother
\begin{document}

\title{On the Deuterium-to-Hydrogen Ratio of the Interstellar Medium}
\author{David H.~Weinberg\altaffilmark{1}}

\altaffiltext{1}{Department of Astronomy and Center for Cosmology
and AstroParticle Physics, The Ohio State University, 
Columbus, OH 43210, dhw@astronomy.ohio-state.edu}

%\begin{abstract}
%\end{abstract}


\begin{abstract}
Observational studies show that the global deuterium-to-hydrogen ratio $\dhr$ in
the local interstellar medium (ISM) is about 90\% of the primordial ratio
predicted by big bang nucleosynthesis. The high $\dhrism$ implies
that only a small fraction of interstellar gas has been processed through 
stars, which destroy any deuterium they are born with. Using analytic arguments
for one-zone chemical evolution models that include accretion and outflow, I
show that the deuterium abundance is tightly coupled to the abundance of core
collapse supernova (CCSN) elements such as oxygen. These models predict that
the ratio of the ISM deuterium abundance to the primordial abundance is 
$\Yd/\Ydp \approx (1+r\Yo/\mocc)^{-1}$, where $r$ is the recycling 
fraction, $\Yo$ is the ISM oxygen mass fraction, and $\mocc$ is the population
averaged CCSN yield of oxygen. Using values $r=0.4$ and $\mocc=0.015$ 
appropriate to a Kroupa (2001) initial mass function and recent CCSN yield
calculations, solar oxygen abundance corresponds to $\Yd/\Ydp\approx 0.87$,
consistent with observations. This approximation is accurate for a wide range
of parameter values, and physical arguments and numerical tests
suggest that it should remain accurate for more complex chemical evolution 
models. 
The good agreement with the upper range of observed $\dhrism$
values supports the long-standing suggestion that sightline-to-sightline 
variations of deuterium are a consequence of dust depletion, rather than a low
global $\dhrism$ enhanced by localized accretion of primordial composition gas.
This agreement limits deviations from conventional yield and recycling values, 
including models in which most high mass stars collapse to form black 
holes without expelling their oxygen in supernovae, 
and it implies that Galactic 
outflows eject ISM hydrogen as efficiently as they eject CCSN metals.  
\end{abstract}

%%%%%%%%%%%%%%%%%%%%%%%%%%%%%% INTRODUCTION %%%%%%%%%%%%%%%%%%%%%%%%%%%%%%%%%
\section{Introduction}
\label{sec:intro}

Most of the atoms of our everyday world were once inside a star.
As astronomers, we are accustomed 
to explaining this remarkable fact in 
our undergraduate classes and popular lectures.
Most atoms in the local interstellar medium (ISM), by contrast,
were never inside a star.  We know this because the
deuterium-to-hydrogen ratio $\dhr$ measured from absorption lines
through the ISM approaches 90\% of the primordial $\dhr$ ratio
\citep{linsky2006}, and stars destroy 100\% of the deuterium
they are born with.\footnote{Proto-stars are fully convective and draw
deuterium nuclei into layers hot enough to fuse them into helium
\citep{bodenheimer1966,mazzitelli1980}.}
Everyday objects are of course biased towards heavy elements,
but even so, this stark difference between the history of atoms
on earth and those in the ISM may seem surprising at first glance.

In fact, as discussed in some detail by \cite{tosi1998},
\cite{romano2006}, and 
\cite{steigman2007}, chemical evolution models that reproduce
other aspects of the local stellar and ISM abundances typically
give good agreement with the observed ISM $\dhr$.  In this paper
I argue that the $\dhr$ ratio should be closely tied to the ISM
oxygen abundance, and that for roughly solar oxygen abundance
the $\dhr$ ratio should be 85-90\% of its primordial value.
This conclusion is consistent with previous findings about
ISM deuterium, and it implies that they are robust to many details of 
the assumed chemical evolution model 
\added{but sensitive to adopted supernova yields.}
My analysis makes use
of an analytic formalism for chemical evolution presented by
%\citeauthor{weinberg2015} (\citeyear{weinberg2015}; hereafter WAF).
Weinberg, Andrews, \& Freudenburg (\citeyear{weinberg2017}, hereafter WAF).
Because this paper focuses on oxygen as a metallicity
tracer, it avoids the more complicated aspects of the WAF formalism
that are connected to SNIa elements.  The modeling approach
can be seen as a generalization of the one originally introduced
by \cite{larson1972}.

The $\dhr$ ratio measured from UV absorption spectroscopy shows
large variations from sightline to sightline through the local
ISM \citep{linsky2006}.  This variation most likely reflects
variable depletion onto dust grains \citep{draine2006,linsky2006},
a point I return to in \S 3.
\cite{linsky2006} estimate a global $\dhr$ in the ISM of
$(2.31 \pm 0.24) \times 10^{-5}$, which is ($90\pm 10$)\% of the value 
$(2.58 \pm 0.13) \times 10^{-5}$
predicted by big bang nucleosynthesis (BBN) for the
baryon density implied by cosmic microwave background 
observations 
(\citealt{cyburt2015}; see also \citealt{nollett2014,coc2015}).
%which is in turn similar to the abundance measured in metal-poor
%quasar absorption systems \citep{cooke2014}.

Before proceeding to a more careful derivation, it is worth giving
an order-of-magnitude argument that explains this result.
For a \cite{kroupa2001} initial mass function (IMF) truncated
at $0.1$ and $100M_\odot$, and
the supernova yields of \cite{chieffi2004}
and \cite{limongi2006}, core collapse supernovae produce
an average of $1.5M_\odot$ of oxygen for every $100 M_\odot$
of star formation (WAF; \citealt{andrews2017}).
%\cite{andrews2015}.  
For the same initial
mass of stars, the total mass of gas returned from SN ejecta 
and the envelopes of AGB stars is $40 M_\odot$ after 2 Gyr and
$45M_\odot$ after 10 Gyr.  If the ISM consisted entirely
of material that had been through stars, then the predicted
oxygen abundance would be approximately
$1.5M_\odot/45M_\odot = 3.3\%$ by mass.
Using the photospheric scale of \cite{lodders2003}, the 
solar oxygen abundance is 0.56\% by mass, 
about 1/6 of this value.
Therefore, if the ISM oxygen abundance is approximately solar,
the material returned from stars must have been diluted by 
five times as much unprocessed gas.\footnote{The 
\cite{lodders2003} solar 
abundance corresponds to $12 + \log({\rm O}/{\rm H}) = 8.69$ on
the conventional shifted number density scale, which is similar
to or slightly above inferred ISM oxygen abundances at the
solar Galactocentric radius (e.g., \citealt{rudolph2006,balser2011}).}
The analysis in the 
next section treats inflow and outflow more explicitly
and leads to the same conclusion.

\section{Evolution of $\ohh$ and $\dhr$}
\label{sec:evolution}

The WAF formalism applies to a one-zone model in which stars
form from a reservoir of gas that is fully mixed at all times.
The star formation rate is $\mdotstar(t)=M_g(t)/\tau_*$ where
$M_g(t)$ is the gas mass at time $t$ and the star formation
efficiency (SFE) timescale $\tau_*$ is assumed constant in time to
allow analytic solutions.  
\added{
Constant $\tau_*$ is equivalent to the ``linear Schmidt law''
adopted in many other analytic models of chemical evolution
(e.g., \citealt{recchi2008}).
}
For the analytic calculations
in this paper I adopt an
exponential star formation history (SFH), 
$\mdotstart = \mdotstaro e^{-t/\tausfh}$, which approaches
a constant star formation rate (SFR) in the limit of long $\tausfh$.
Star formation is assumed to drive outflows with a constant
mass-loading factor $\eta = \mdotout/\mdotstar$.  
\added{
I use numerical calculations to examine some more complex
scenarios in \S\ref{sec:numerical}.
}

\subsection{Oxygen}

In the instantaneous recycling approximation,
the evolution equation for the total mass of oxygen in the ISM is
\begin{equation}
\label{eqn:MoDiffEq}
\Modot = \mocc\mdotstar - \Yo\mdotstar - \Yo\eta\mdotstar + \Yo r\mdotstar 
\end{equation}
where $\Yo(t)$ is the current ISM oxygen abundance by mass.
The first term represents enrichment by core collapse supernovae
(CCSNe) with an IMF-averaged yield of $\mocc$, i.e., for each solar
mass of star formation $\mocc$ solar masses of oxygen are produced
by massive stars and returned to the ISM.  I adopt $\mocc=0.015$
based on the IMF and yield assumptions described in \S\ref{sec:intro}.
The second term represents depletion of oxygen already in the ISM
into stars.  The third term represents depletion by outflow, with
mass-loading factor $\eta$.  Producing solar abundances 
at late times requires $\eta \approx 2.5$ (WAF).
The fourth term represents recycling of the oxygen stars were
originally born with, where $r$ is the fraction of mass formed
into stars that is returned to the ISM by supernovae and by the 
winds of evolved stars.  For a \cite{kroupa2001} IMF the recycled
fraction is $r(t) = 0.37$, 0.40, and 0.45 after 1, 2, and 10 Gyr,
respectively.  The approximation in equation~(\ref{eqn:MoDiffEq})
treats this recycling as instantaneous with a single effective
value of $r$, and WAF show that for $r=0.4$ this approximation
reproduces numerical results with full time-dependent recycling quite
accurately.  

The gas mass in the ISM is depleted by star formation and outflow
and replenished by recycling and infall, with time derivative
\begin{equation}
\label{eqn:MgDeriv}
\dot{M}_g(t) = -(1+\eta-r)\mdotstar+\mdotinf(t)~.
\end{equation}
In the absence of accretion, star formation and outflow with
$\mdotstart = M_g(t)/\taustar$ would 
deplete the gas supply on an $e$-folding timescale 
\begin{equation}
\label{eqn:taudep}
\taudep = \taustar/(1+\eta-r)~.
\end{equation}
More generally, 
the infall rate $\mdotinf(t)$ is determined implicitly by the adopted SFE
timescale and star formation history.
For constant $\taustar$ one can set $\dot{M}_g = \taustar\ddot{M}_*$, and
with an exponential SFH yielding $\ddot{M_*} = -\mdotstar/\tausfh$ 
one obtains
\begin{equation}
\label{eqn:mdotinf}
\mdotinf = (1+\eta-r-\taustar/\tausfh)\mdotstar~.
\end{equation}
I assume below that infalling gas has primordial composition, i.e., 
no oxygen and the BBN value of $\dhr$.

For an exponential SFH,
equation~(\ref{eqn:MoDiffEq}) can be rewritten
\begin{equation}
\label{eqn:MoDiffEq2}
\Modot + {\Mo\over\taudep} = \mocc\mdotstaro e^{-t/\tausfh}~.
\end{equation}
One can apply a standard solution method for this type of differential equation,
\begin{equation}
\label{eqn:StandardSolution}
\Mo(t) = {1\over \mu(t)}\left[\int_0^t\mu(t')f(t')dt'\right]~,
\end{equation}
where $\mu(t) = e^{t/\taudep}$, $f(t)$ is the driving term on the
r.h.s.\ of equation~(\ref{eqn:MoDiffEq2}), and an arbitrary
integration constant has been set to zero to satisfy 
the boundary condition $\Mo=0$ at $t=0$.
The result is
\begin{equation}
\label{eqn:MoEvol}
\Mo(t) = \mocc\mdotstaro\taubar\left(e^{-t/\tausfh}-e^{-t/\taudep}\right)~,
\end{equation}
where I have introduced the notation
\begin{equation}
\label{eqn:taubar}
\taubar = \left({1\over \taudep}-{1\over\tausfh}\right)^{-1}
\end{equation}
for a ``harmonic difference timescale'' composed of the gas depletion
and SFH timescales.  (In WAF, where there are several such timescales,
this particular quantity is denoted $\taubarDepSfh$.)
To get the oxygen abundance, one divides $\Mo(t)$ by
$M_g(t) = \taustar \mdotstart$, and with the notation~(\ref{eqn:taubar})
one can express the result as
\begin{equation}
\label{eqn:YoEvol}
\begin{split}
\Yo(t) &= \mocc {\taubar\over\taustar}\left(1-e^{-t/\taubar}\right) \\
       &= \Yoeq\left(1-e^{-t/\taubar}\right)~.
\end{split}
\end{equation}
Here $\Yoeq$ is the ``equilibrium'' oxygen abundance approached at times
$t \gg \taubar$:
\begin{equation}
\label{eqn:YoEq}
\Yoeq = \mocc {\taubar\over\taustar} = {\mocc \over 1+\eta-r-\taustar/\tausfh}~.
\end{equation}
Once the ISM reaches this equilibrium abundance, enrichment from further
star formation is balanced by dilution from infall and depletion by
star formation and outflow.

\subsection{Deuterium}

The deuterium mass in the ISM changes because of accretion, which brings
in gas at the primordial deuterium abundance, and because of star formation
and outflows, which consume gas at the current ISM abundance.
The evolution equation\footnote{Note that $\Yd$ and $\Ydp$ refer to the
deuterium mass fraction, while $\dhr$ and $\dhrp$ refer to 
the deuterium-to-hydrogen ratio by number.  The ratios
$\Yd/\Ydp$ and $\dhr/\dhrp$ are, of course, equal.}
is
\begin{equation}
\label{eqn:MdDiffEq}
\Mddot = \Ydp\mdotinf - \Yd(1+\eta)\mdotstar~.
\end{equation}
In contrast to oxygen evolution, there is no recycling term
because any deuterium that enters a star is destroyed before
being returned to the ISM.  
Using $\Yd\mdotstar = \Yd M_g/\taustar = \Md/\taustar$ and
equation~(\ref{eqn:mdotinf}) for the infall rate with an exponential SFH
yields
\begin{equation}
\label{eqn:MdDiffEq2}
\Mddot + {1+\eta\over\taustar}\Md = \Ydp (1+\eta-r-\taustar/\tausfh)
   \mdotstaro e^{-t/\tausfh}~.
\end{equation}
This equation can be solved by the same technique used previously
for oxygen, with the boundary condition that $\Md = \Ydp M_g$ at $t=0$.
After some manipulation, the result can be expressed in the form
\begin{equation}
\label{eqn:YdEvol}
\Yd(t) = \Ydp (1+C)^{-1}\left[1+C e^{-rt/\taustar} e^{-t/\taubar}\right]
\end{equation}
with the constant
\begin{equation}
\label{eqn:Cdef}
C = {r \over 1+\eta-r-\taustar/\tausfh} = r {\Yoeq\over \mocc}~,
\end{equation}
where $\Yoeq$ is the equilibrium oxygen abundance of
equation~(\ref{eqn:YoEq}).
For $r=0$, equation~(\ref{eqn:YdEvol}) yields
$\Yd=\Ydp$ at all times, which is as expected because in this
case all hydrogen in the ISM is primordial by definition.
The solution~(\ref{eqn:YdEvol}) can be verified by direct
substitution into equation~(\ref{eqn:MdDiffEq2}), noting 
that the r.h.s.\ of~(\ref{eqn:MdDiffEq2}) can be written 
as $\Ydp\mdotstart\times (r/C)$.

At $t=0$, equation~(\ref{eqn:YdEvol}) yields $\Yd=\Ydp$ as expected.
However, once $t$ becomes large compared to either $\taustar/r$
or $\taubar$, then $\Yd$ approaches an equilibrium value
\begin{equation}
\label{eqn:YdEq}
\Ydeq = \Ydp (1+C)^{-1} = {\Ydp \over 1+r\Yoeq/\mocc}~.
\end{equation}
Continuing infall leads to an equilibrium in which the loss
of deuterium to star formation and outflow is balanced by 
accretion of primordial composition gas.
Equation~(\ref{eqn:YdEq}) is the principal mathematical result
of this paper, showing the relation between ISM deuterium
and oxygen abundances in equilibrium, and the dependence of 
this relation on the physical parameters $\Ydp$, $r$, and $\mocc$.

\subsection{Evolutionary Tracks}

%%%%%%
\begin{figure}
\centerline{\includegraphics[width=2.8truein]{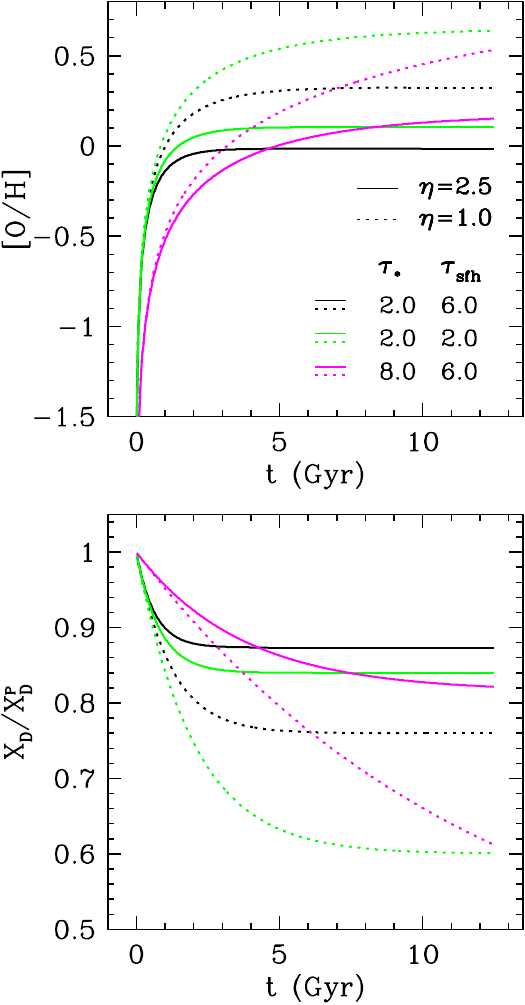}}
\caption{Evolution of $\ohh = \log_{10} \Yo/Z_{{\rm O},\odot}$ (top)
and the deuterium abundance (bottom, scaled to the primordial value)
for a selection of one-zone models.
Solid curves represent models with outflow mass loading factor 
$\eta=2.5$, while dotted curves adopt a lower $\eta=1.0$ that
leads to super-solar oxygen abundance.
Black curves show a typical SFE timescale ($\taustar=2\Gyr$)
and slowly declining SFH ($\tausfh=6\Gyr$).  Green curves show the
same SFE but a rapidly falling SFH ($\tausfh=2\Gyr$).  Magenta
curves show a model with inefficient star formation ($\taustar=8\Gyr$)
and slowly declining SFH ($\tausfh=6\Gyr$).
}
\label{fig:evol}
\end{figure}
%%%%%%

Figure~\ref{fig:evol} plots the evolution of the oxygen and
deuterium abundances for a variety of parameter choices.
The solid black curves shows a case with an SFE timescale
$\taustar=2\Gyr$ typical of that found for molecular gas
in star-forming galaxies \citep{leroy2008}, a slowly declining
star formation history with $\tausfh=6\Gyr$, and a 
mass-loading parameter $\eta=2.5$ chosen to yield $\Yoeq$
near solar.  With a depletion timescale of $0.65\Gyr$ (eq.~\ref{eqn:taudep}),
the oxygen abundance rises quickly to its equilibrium value, and
deuterium declines to the corresponding equilibrium within $2\Gyr$.
The solid green curves have $\tausfh=2\Gyr$, which raises $\Yoeq$
because of the more rapidly declining gas supply.
The equilibrium deuterium abundance is correspondingly lower.
Magenta solid curves show a model with much lower star formation
efficiency, $\taustar=8\Gyr$.  Here the approach to equilibrium
is much slower, though both oxygen and deuterium are nearly
constant after $t=10\Gyr$.  

Dotted curves show the same cases but with a lower mass-loading
factor $\eta=1.0$.  Here the equilibrium oxygen abundances are
higher because more of the metals produced by stars are retained,
and the equilibrium deuterium abundances are lower because
a larger fraction of the ISM consists of stellar ejecta.
The lower outflow rate also lengthens the gas depletion time,
and for $\taustar=8\Gyr$ the abundances have not yet reached
their equilibrium values by $t=12.5\Gyr$.
\added{
For all of these models, the predicted trend of $\ohh$ with time
is nearly flat over the past $8-10\Gyr$, in qualitative 
agreement with the observed age-metallicity relation of stars
(e.g., \citealt{edvardsson1993,bensby2014,feuillet2016}),
but explaining the detailed trends and the considerable scatter
in the age-metallicity relation requires radial mixing
of stellar populations, beyond the scope of the one-zone models
used here (D.\ Feuillet et al., in prep.).
}

%%%%%%
\begin{figure}
\centerline{\includegraphics[width=2.8truein]{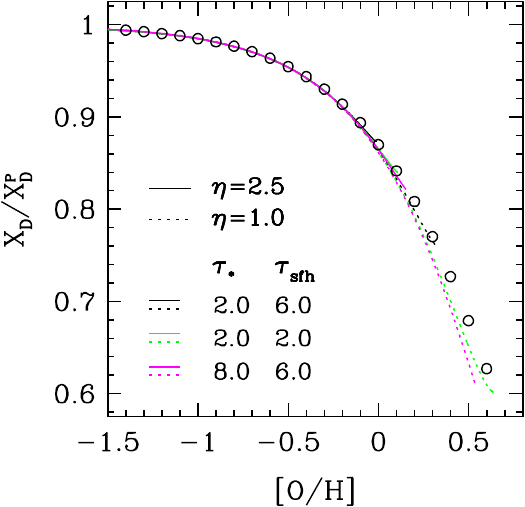}}
\caption{Deuterium abundance, scaled to the primordial value, as
a function of oxygen abundance $\ohh$, for the models plotted
in Fig.~\ref{fig:evol}.  Circles show the approximate formula
$\Yd/\Ydp = 1/(1+r\Yo/\mocc)$, where $r$ is the recycling fraction
($r=0.4$ in these models), $\Yo$ is the ISM oxygen abundance by mass,
and $\mocc$ is the IMF-averaged oxygen yield of CCSNe.
The models represented by the green and magenta dotted curves 
have not reached equilibrium by $t=12.5\Gyr$; if they are extended
further in time then they converge to the locus shown by the circles.
}
\label{fig:track}
\end{figure}
%%%%%%

Figure~\ref{fig:track} plots the evolution of these models in
the plane of $\Yd/\Ydp$ vs. $\ohh$.  Remarkably, they are nearly
superposed on a universal curve.  Circles show the obvious
generalization of equation~(\ref{eqn:YdEq}),
\begin{equation}
\label{eqn:YdApprox}
\Yd = {\Ydp \over 1+r\Yo/\mocc}~,
\end{equation}
which simply assumes that the equilibrium relation also
applies at earlier times.  This formula
describes the model results very well, though it slightly
overpredicts $\Yd$ for super-solar $\ohh$.
I have tried other values of $r$ and $\mocc$ and find
equally good agreement with equation~(\ref{eqn:YdApprox}).

Why does a formula derived for equilibrium abundances hold so well 
for systems that have not yet reached equilibrium?
The mathematical answer is not obvious, 
but the ansatz of equation~(\ref{eqn:YdApprox})
is closely connected to the approximate argument given in the
introduction.  Suppose that the ISM consisted of a mix of
primordial gas, with zero oxygen abundance,
and gas that had been through stars precisely once,
with oxygen abundance $\mocc/r$.
The overall ISM oxygen abundance in this case would be
$\Yo = \frec\mocc/r = (1-\fprim)\mocc/r$, where $\frec$
and $\fprim = 1-\frec$ are the mass fractions of recycled and primordial
material, respectively.
The deuterium abundance is $\Ydp\fprim$, and substituting
for $\fprim$ gives 
\begin{equation}
\label{eqn:accounting}
\Yd \approx \Ydp(1-r\Yo/\mocc)~,
\end{equation}
which is equal
to equation~(\ref{eqn:YdApprox}) at first order in $r\Yo/\mocc$.
Equation~(\ref{eqn:accounting}) is approximate because some 
gas in the recycled component has been through stars more than
once, making the oxygen abundance in this component higher
than $\mocc/r$, and it is 
somewhat less accurate than equation~(\ref{eqn:YdApprox}) at
super-solar $\ohh$.
Nonetheless, the accuracy of equation~(\ref{eqn:YdApprox}) appears
to be rooted in fairly basic accounting,
and it becomes exact in the limit of equilibrium.

\subsection{Metal-Enriched Winds}
\label{sec:enhanced}

Equation~(\ref{eqn:MoDiffEq}) implicitly assumes that the gas
ejected in winds has the current ISM abundance $\Yo(t)$.
If winds are driven by energy or momentum input from supernovae 
and massive stars, then it is possible that the metallicity
of ejected material is higher than that of the ambient ISM,
\added{
with significant impact on chemical evolution predictions
(e.g., \citealt{pilyugin1993,marconi1994,recchi2008,spitoni2015}).
}
In the extreme limit, CCSN ejecta could escape from the galaxy
without entraining ISM gas at all.  The relation between ISM
oxygen and deuterium abundances can set constraints on the
degree of metal-enhancement in winds.

If the metallicity (specifically, the oxygen abundance)
of ejected material is higher than that of ambient ISM gas by
a factor 
\begin{equation}
\label{eqn:xi_enh}
\xi_{\rm enh} = {Z_{\rm wind} \over Z_{\rm ism}}~,
\end{equation}
then the third term on the r.h.s. of equation~(\ref{eqn:MoDiffEq})
is multiplied by $\xi_{\rm enh}$ while other terms are unchanged.
The solution for oxygen evolution is the same as before except
that $\eta$ is replaced by $\eta\xi_{\rm enh}$, the effective
mass-loading factor for oxygen ejection, in both the equilibrium
abundance (eq.~\ref{eqn:YoEq}) and the depletion 
timescale~(eq.~\ref{eqn:taudep}).
Thus, for $\xi_{\rm enh}>1$ and a given $\eta$, oxygen evolves to a lower
equilibrium abundance and reaches that equilibrium more quickly.

Deuterium evolution depends on the overall mass outflow rate,
independent of $\xi_{\rm enh}$, so equation~(\ref{eqn:YdEvol})
and the first equality of equation~(\ref{eqn:Cdef}) are unchanged.
However, metal-enhanced winds alter the relation between
the equilibrium deuterium and oxygen abundances to
\begin{equation}
\label{eqn:YdEqEnh}
\Ydeq = {\Ydp \over 1+\beta r\Yoeq'/\mocc}
\end{equation}
where $\Yoeq'$ represents the equilibrium oxygen 
abundance in the metal-enhanced case and 
\begin{equation}
\label{eqn:beta}
\beta = {1 + \eta\xi_{\rm enh} - r -\taustar/\tausfh \over
         1+\eta-r-\taustar/\tausfh} ~.
\end{equation}
For $\xi_{\rm enh}>1$, the deuterium abundance at a given 
oxygen abundance is lower. 
Physically, this lower $\Ydeq$ arises
because the mass loading factor $\eta$
required to produce the specified oxygen abundance is lower, 
and lower outflow implies correspondingly lower accretion at fixed SFR.
Reduced accretion in turn implies that a larger fraction of the ISM
consists of recycled stellar material, depleted in deuterium.

\begin{figure}
\centerline{\includegraphics[width=2.5truein]{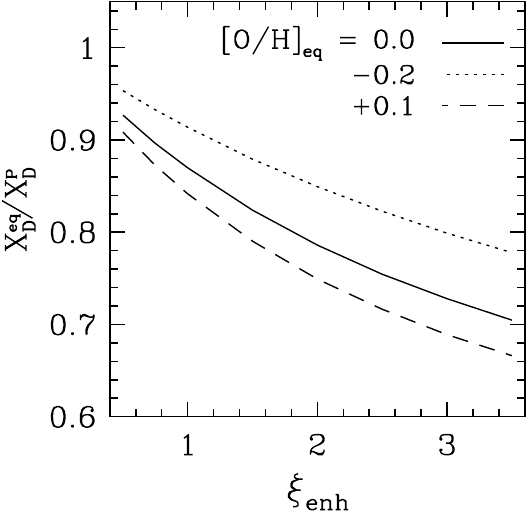}}
\caption{Ratio of equilibrium deuterium abundance to the 
primordial abundance as a function of the enhancement
factor $\xi_{\rm enh} = Z_{\rm wind}/Z_{\rm ISM}$, the
ratio of wind metallicity to ISM metallicity.
Solid, dotted, and dashed curves show cases in which the
equilibrium oxygen abundance is solar, $\ohh=-0.2$,
and $\ohh=+0.1$, respectively.  All cases assume $\taustar/\tausfh=1/3$.
}
\label{fig:enhanced}
\end{figure}

Figure~\ref{fig:enhanced} illustrates the magnitude of this
effect, where I have adopted $\taustar=2.0\Gyr$ and $\tausfh=6.0\Gyr$.
For each value of $\xi_{\rm enh}$, the value of $\eta$ is
chosen to yield a solar equilibrium abundance (solid curve)
or $\ohh_{\rm eq} = -0.2$ (dotted curve) or +0.1 (dashed curve).
The equilibrium deuterium abundance decreases with increasing $\xi_{\rm enh}$,
as implied by equation~(\ref{eqn:YdEqEnh}).
For example, achieving solar oxygen abundance in the standard case
of $Z_{\rm wind} = Z_{\rm ISM}$ requires $\eta = 2.4$, with
an equilibrium $\Ydeq/\Ydp = 0.87$.  However, with a metal-enhancement factor
$\xi_{\rm enh}=2$, the required value of $\eta$ is 1.2, and the
implied $\Ydeq/\Ydp = 0.79$.
Unfortunately, dust depletion uncertainties make precise determinations
of $\dhrism$ challenging, but this value is probably at the lower
limit of the acceptable range from \cite{linsky2006}.
Metal enhancement with $\xi_{\rm enh} \geq 3$, implying
$\Ydeq/\Ydp \leq 0.73$, would be difficult to reconcile
with the observed deuterium abundance.
There is no obvious physical mechanism for producing metal-depleted
outflows ($\xi_{\rm enh} < 1$), but if they occurred they would
lead to higher $\Yd$ at a given $\ohh$.

%%%%%%
%%%%%%

\subsection{Numerical Examples}
\label{sec:numerical}

\begin{figure*}
\centerline{\includegraphics[width=5.5truein]{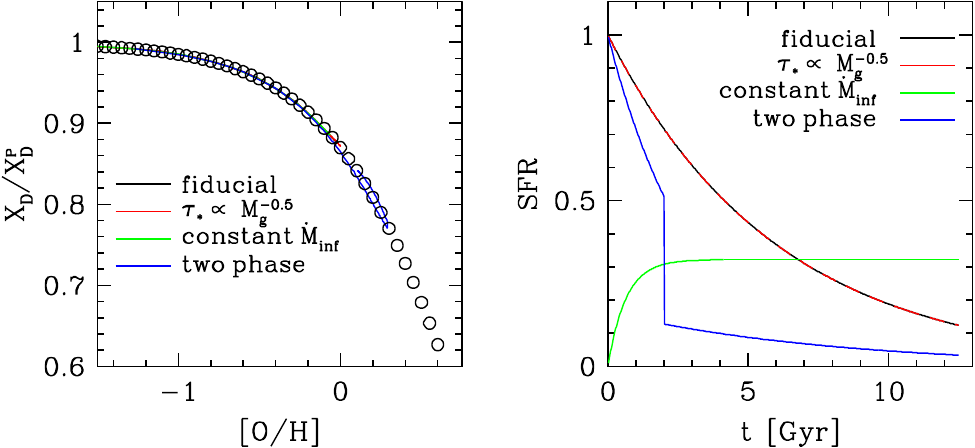}}
\caption{Evolutionary tracks of $\Yd/\Ydp$ vs.\ $\ohh$ (left) for 
models with star formation histories shown in the
right panel.  Black curves show our fiducial exponential model
with constant $\taustar$, red curves a model with the same
star formation history and ``Schmidt Law''
$\taustar$ scaling, green curves a model with constant mass infall
rate, and blue curves a model in which star formation parameters
and $\eta$ change suddenly at $t=2\Gyr$ (see text).
Evolutionary tracks are nearly indistinguishable and agree
well with equation~(\ref{eqn:YdApprox}) shown by open circles,
but one can see that the two phase model loops back to lower $\ohh$
and higher $\Yd/\Ydp$ after $\eta$ changes from 1 to 2.
}
\label{fig:models}
\end{figure*}

The analytic solutions here apply to a restricted class of models,
so to investigate other cases I have written a simple numerical
code that integrates the governing differential 
equations~(\ref{eqn:MoDiffEq}) and~(\ref{eqn:MdDiffEq}).
Figure~\ref{fig:models} compares three different models to
a fiducial analytic model with constant SFE timescale $\taustar=2\Gyr$,
an exponential star formation history with $\tausfh=6\Gyr$,
and $\eta=2.5$.  Observed star formation laws
\citep{schmidt1959,kennicutt1998} show a non-linear relation between
SFR surface density and total gas surface density, approximately
$\Sigma_{\rm SFR} \propto \Sigma_g^{1.5}$.  If we think of 
a one-zone chemical evolution model as representing an annulus
of the Galactic disk, we might therefore expect the SFE timescale
to grow as the gas surface density decreases, with 
$\taustar \propto M_g^{-0.5}$.  Red curves show a numerical 
model with this SFE scaling and the same exponential star formation
history.  The predicted evolutionary track of 
$\Yd/\Ydp$ vs.\ $\ohh$ is indistinguishable from that of the
constant $\taustar$ model and in excellent agreement with
equation~(\ref{eqn:YdApprox}).  

%%%%%%
%%%%%%

Green curves show a model with constant $\taustar=2\Gyr$ and a constant
mass infall rate of $1 M_\odot\yr^{-1}$, again with $\eta=2.5$.
With zero initial gas mass, the star formation rate of this model
rises linearly at early times and asymptotes to an equilibrium 
value of $\mdotstar = \mdotinf/(1+\eta-r)$.  Despite the very
different star formation history (as shown in the right panel),
the track of this model in the
$\Yd$-$\ohh$ plane is indistinguishable from those of the 
exponential models.  
\added{
A linear-exponential star formation history,
similar in form to that modeled by
Spitoni et al.\ (\citeyear{spitoni2017}; see Appendix B of WAF for
the relation between these models),
also yields a very similar result (not shown).}
Blue curves show a model with a sharp transition
of parameters at $t=2\Gyr$, from efficient early star
formation with low outflow mass loading
($\taustar=1\Gyr$, $\tausfh=3\Gyr$, $\eta=1$) 
to slower star formation and higher outflow efficiency
($\taustar=4\Gyr$, $\tausfh=8\Gyr$, $\eta=2$) at later times.
Because of the low initial $\eta$,
this model first evolves to super-solar $\ohh$ and $\Yd/\Ydp \approx 0.78$.
After the transition to higher $\eta$, it loops back to a new 
equilibrium at lower $\ohh$ and higher $\Yd/\Ydp$, approximately but
not exactly reversing its original evolutionary track.

Although these examples are not exhaustive, they suggest that the
behavior of equation~(\ref{eqn:YdApprox}) applies to a wide range
of models, provided that outflows eject gas at the ISM metallicity.
\added{
Further support for this view comes 
\cite{vandevoort2017}, who investigate deuterium evolution in
3-d cosmological hydrodynamic simulations of galaxy formation,
which automatically lead to complex outflows and gas mixing processes
and produce gas with a wide range of oxygen and deuterium abundances.
They find that the median relation between $\Yd/\Ydp$ and $\ohh$ 
is in excellent agreement with equation~(\ref{eqn:YdApprox}),
and that the $1\sigma$ scatter about this median relation 
is small.
}

\subsection{Supernova Yields and the IMF}
\label{sec:bhcut}

\begin{figure}
\centerline{\includegraphics[width=3.2truein]{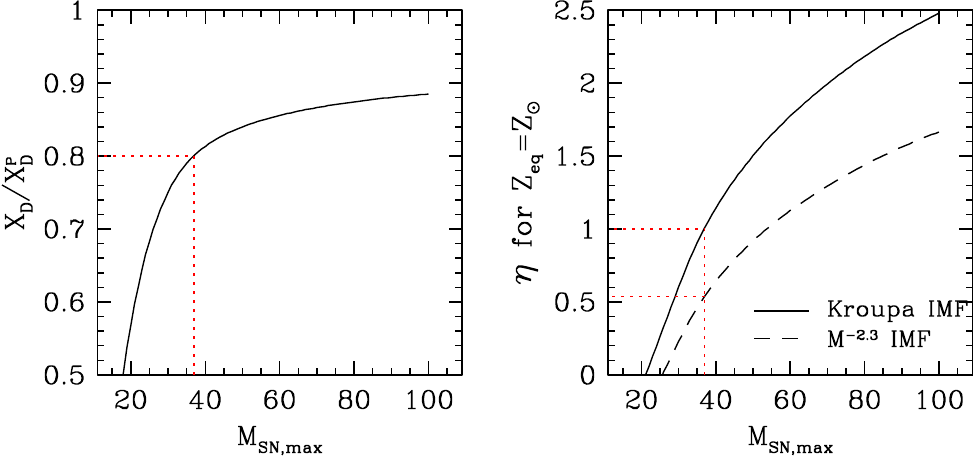}}
\caption{
{\it (Left)} Influence of the upper mass cutoff for supernovae on
the predicted deuterium abundance at solar metallicity,
computed for a \cite{kroupa2001} IMF with the oxygen yields
of \cite{limongi2006}.  A value $M_{\rm SN,max}=37 M_\odot$
yields $\Yd/\Ydp=0.8$, the $1\sigma$ lower bound on the
ISM deuterium abundance from \cite{linsky2006}.
{\it (Right)} The value of $\eta$ required to yield solar oxygen
abundance at equilibrium, as a function of $M_{\rm SN,max}$.
The solid curve shows results for a \cite{kroupa2001} IMF and
the dashed curve for an $M^{-2.3}$ power-law IMF that extends
down to $0.1 M_\odot$.  A $37 M_\odot$ supernova cutoff 
corresponds to $\eta = 1.0$ or 0.54 in these two cases.
Changing the IMF at low masses (below the main sequence turnoff)
does not influence the predicted $Y_D$ because
it has canceling effects on $\mocc$ and $r$.
}
\label{fig:bhcut}
\end{figure}

The key parameters influencing the relation between oxygen abundance
and deuterium abundance are the oxygen yield $\mocc$ and the 
recycling fraction $r$.  For a \cite{kroupa2001} IMF truncated
at $0.1M_\odot$ and $100 M_\odot$, the solar metallicity supernova yields 
of \cite{limongi2006} yield $\mocc = 0.017$ assuming that all
stars above $8M_\odot$ explode as CCSNe (see \citealt{andrews2017}
for details of the IMF-averaged yield calculation).
However, the predicted oxygen production is a steeply increasing
function of progenitor mass, and if the most massive stars
collapse to form black holes instead of exploding then the
IMF-averaged yield can be significantly reduced.
The left panel of
Figure~\ref{fig:bhcut} shows the predicted value of $\Yd/\Ydp$
at solar $\ohh$, computed from equation~(\ref{eqn:YdApprox}),
assuming that all stars with $M > M_{\rm SN,max}$ collapse to
form black holes and release no oxygen.\footnote{I am grateful
to Brett Andrews for providing me with IMF-averaged yields as
a function of $M_{\rm SN,max}$, calculated using the \cite{limongi2006}
supernova yield tables.}  I have used the recycling fraction 
$r=0.4$, appropriate for a \cite{kroupa2001} IMF after 2 Gyr,
and I have ignored the small impact of black hole formation on $r$
under the assumption that massive stars still return most of the
mass they were born with.  For a cutoff $M_{\rm SN,max} < 37 M_\odot$
the predicted $\Yd/\Ydp$ falls below 0.8, in tension with the
\cite{linsky2006} ISM estimates.

%%%%%%
%%%%%%

The right panel of Figure~\ref{fig:bhcut} shows the value of $\eta$
required to produce $\Yoeq=Z_{\rm O,\odot}$,
\begin{equation}
\eta = {\mocc \over Z_{\rm O,\odot}} + r -1 - \taustar/\tausfh ~,
\label{eqn:etareq}
\end{equation}
assuming $r=0.4$ and $\taustar/\tausfh = 1/3$ as in the fiducial model.
As $\mocc$ decreases with decreasing $M_{\rm SN,max}$, the required
value of $\eta$ also drops.  However, values of $M_{\rm SN,max}$
that imply $\eta \la 1$ also imply $\Yd/\Ydp \la 0.8$, so one
cannot eliminate the need for outflows by reducing the oxygen yield
without running afoul of the deuterium constraint.

The \cite{kroupa2001} IMF scales as $M^{-2.3}$ for $M > 0.5M_\odot$
and as $M^{-1.3}$ for $M = 0.1-0.5 M_\odot$.  A pure
$M^{-2.3}$ power law IMF has more low mass stars, 
increasing the total mass of the stellar population 
by a factor $\approx 1.3$ for the same number of high mass stars.
(A Salpeter [\citeyear{salpeter1955}] IMF has a slightly steeper,
$M^{-2.35}$ slope.)
The dashed curve in Figure~\ref{fig:bhcut}
shows the value of $\eta$ implied by
equation~(\ref{eqn:etareq}) after reducing both $\mocc$ and $r$
by this factor of 1.3.  Adding low mass stars does not change
the $\Yd/\Ydp$ results shown in the left panel because 
the changes to $\mocc$ and $r$ cancel in equation~(\ref{eqn:YdApprox}).
A bottom-heavy IMF reduces the level of outflow required to
achieve a solar ISM oxygen abundance, but reducing $\eta$ to zero
is still not possible without an observationally unacceptable
level of deuterium depletion.  A bottom-heavy IMF of this sort
also conflicts with the observed stellar populations of the
local disk \citep{kroupa2001,chabrier2003} and with the
dynamical mass-to-light ratios of spiral galaxies \citep{bell2001}.

\added{
Many models of the galaxy mass-metallicity relation find that
strong outflows are required to reproduced observed ISM
metallicities (e.g., \citealt{finlator2008,peeples2011,zahid2012}),
as implied by Figure~\ref{fig:bhcut}.  However, some Galactic 
chemical evolution models reproduce properties of the solar 
neighborhood and the Milky Way disk without invoking outflows.
Older papers often treated population-averaged yields as effectively
free parameters, and some that did use computed supernova yields
overpredicted the solar neighborhood abundance by a factor 
of two or more even for a Salpeter IMF (e.g., \citealt{matteucci1986}).
Radial gas flows can mitigate the need for outflows by driving
enriched gas from the solar annulus inward and replacing it
with lower metallicity gas from the outer Galaxy.
Fountains of ejected gas that fall back to the disk
can also mimic the impact of permanent outflows provided
that returning gas is diluted by enough entrained primordial
material to reduce its metallicity by a substantial factor.
Some no-outflow models with and without radial gas flows yield
reasonable agreement with observed ISM oxygen abundances
\citep{molla2005,spitoni2011,cavichia2014}.  It is difficult
to tell how much of the difference from one-zone models is a
consequence of yields, since most papers detail their calculations
but do not summarize them in a form (such as the $\mocc$ parameter
given here) that enables straightforward comparison.
Given the arguments presented here, it would be especially interesting
to investigate the deuterium predictions of no-outflow models
that reproduce the observed ISM oxygen abundance.
}

\section{Implications}
\label{sec:implications}

The results in \S\ref{sec:evolution} agree with those of more sophisticated
models of galactic chemical evolution (GCE), which generally yield ISM
deuterium abundances only moderately below the primordial abundance
when they are tuned to reproduce observations of the solar neighborhood
(e.g., \citealt{tosi1998,romano2006,steigman2007,lagarde2012}).
The simplicity of the arguments leading to equations~(\ref{eqn:YdEq})
and~(\ref{eqn:YdApprox}) implies that this behavior should be robust 
to details of the GCE model, and it shows that the most important
assumptions are likely those that affect the oxygen yield $\mocc$ and
the recycling fraction $r$.  The numerical examples illustrated in
Figure~\ref{fig:models} 
\added{
and the hydrodynamic simulation results of \cite{vandevoort2017}}
provide further support for 
the robustness of this prediction.
A mixture of gas from zones that
individually obey equation~(\ref{eqn:YdApprox}) will also follow
equation~(\ref{eqn:YdApprox}) if $r\Yo/\mocc \ll 1$ (allowing a
linear Taylor expansion) and the 
level of deuterium depletion is therefore small in each zone.
However, a mixture of high-metallicity, heavily depleted gas 
with low-metallicity, low depletion gas would initially lie 
above this locus (high deuterium abundance relative to $\ohh$),
then return to it as further star formation drove the system
toward equilibrium.
It is thus likely but not guaranteed 
that equation~(\ref{eqn:YdApprox}) will remain accurate
in more sophisticated GCE models that incorporate ingredients such as
radial gas flows and fountain recycling.

The close links predicted between oxygen and deuterium abundances
strengthen arguments \citep{draine2006,linsky2006,steigman2007} that 
sightline-to-sightline variations of $\dhr$ absorption are caused
by depletion of deuterium onto dust, as originally proposed 
by \cite{jura1982}.  This explanation requires a large fraction of the 
available sites on polycyclic aromatic hydrocarbon (PAH) molecules
to be occupied by deuterium rather than hydrogen, but the energy
differences associated with the heavier deuterium nucleus permit
this preferential outcome in the cool ISM, and reaction rates
appear high enough to achieve it \citep{draine2006}.  
Producing factor-of-two variations in $\dhr$ by differential astration,
on the other hand,
would induce enormous variations in $({\rm O}/{\rm H})$, since it would
require a much larger fraction of the ISM on low $\dhr$ sightlines
to be comprised of stellar ejecta.  The alternative of setting
the mean ISM $\dhr$ to $\sim 1-1.5\times 10^{-5}$ and producing
high $\dhr$ by localized infall of primordial gas would require a major 
revision of oxygen yields to avoid overproducing oxygen.  
With our adopted $\mocc = 0.015$, reaching $\dhr=1.0\times 10^{-5}$
implies $\ohh=+1.0$, and $\dhr=1.5\times 10^{-5}$ implies $\ohh=+0.7$.

Many studies have used $\dhr$ in low metallicity, extragalactic
Lyman-$\alpha$ absorption systems to estimate the primordial $\dhr$
ratio and thereby constrain the mean baryon density
(e.g., \citealt{cooke2014} and references therein).  
The baryon density inferred
from observations of the cosmic microwave background
\citep{hinshaw2013,PlanckXIII2015} agrees well with that
inferred from $\dhr$ observations, an important cosmological
consistency test that constrains non-standard BBN models
\citep{cyburt2015,nollett2015}.
Equation~(\ref{eqn:YdApprox}) implies that evolutionary 
corrections to $\dhr$ should be small in systems with sub-solar
oxygen abundance, and there is no need to seek out ultra-low
metallicity systems to eliminate these corrections.  The primary
reason to concentrate on low metallicity sightlines for
cosmological $\dhr$ studies is to reduce uncertainties 
associated with dust depletion.
\cite{dvorkin2016}, using more sophisticated chemical evolution
models based on cosmologically motivated infall rates,
reach similar conclusions about the small
impact of deuterium depletion in damped Lyman-$\alpha$ systems
and the tight expected correlation between oxygen and deuterium
abundance.

Like deuterium, $\hethree$ is produced in BBN at fairly
high abundance, $\hehp \approx 10^{-5}$ \citep{cyburt2015}.
In contrast to deuterium, $\hethree$ can be produced in
stars as well as destroyed \citep{iben1967,truran1971,rood1976,dearborn1996}, 
so even the sign of evolutionary corrections is not
obvious without detailed modeling.
The arguments presented here suggest that the magnitude
of $\hethree$ destruction should be small because most ISM
gas has not been processed by stars at all.
Observations indicate an ISM $\hethree$ abundance that is 
not far from the BBN value \citep{gloeckler1996,bania2002}.
In concert with minimal destruction, this small amount of evolution
implies that production of $\hethree$ by stellar nucleosynthesis
must be small.  As discussed in detail by \cite{lagarde2012},
this result is a challenge to conventional stellar evolution
models but is well explained by models that incorporate
thermohaline mixing.

The robustness of the $\dhr$ prediction to detailed
assumptions implies that deuterium observations provide only limited
constraints on standard GCE models; most models that produce solar
$\ohh$ should predict $\dhr$ that matches current observations
within their uncertainties.  
However, if dust depletion uncertainties
can be controlled, then precise $\dhr$ measurements could
provide a useful test of yield and recycling values, especially
if deuterium can be mapped over a wide enough range of metallicity
to demonstrate the behavior predicted by equation~(\ref{eqn:YdApprox}).
As shown in Figure~\ref{fig:bhcut}, the constraint $\Yd/\Ydp \geq 0.8$
is already enough to challenge otherwise plausible scenarios
in which most stars above $30 M_\odot$ collapse to black holes
instead of releasing oxygen in supernovae
(see discussions by, e.g., \citealt{pejcha2015,sukhbold2016}).

As discussed in \S\ref{sec:enhanced}, the prediction 
of equations~(\ref{eqn:YdEq}) and~(\ref{eqn:YdApprox}) 
is violated in models where outflows preferentially drive
out CCSN ejecta while retaining the AGB ejecta returned 
over longer timescales.
This scenario allows more return of deuterium-depleted hydrogen
for a given amount of oxygen, so it predicts lower $\dhr$ as 
a function of $\ohh$.  The agreement of observed abundances
with simple predictions disfavors metal-enhanced winds with 
$\xi_{\rm enh} \ga 2$.
Models in which radiation pressure ejects galactic dust
with minimal gas entrainment (e.g., \citealt{aguirre2001})
could also violate $\dhr$ constraints, though here one must take
account of the potentially high fraction of deuterium residing
in ejected dust.

The analytic models presented here adopt the instantaneous recycling
approximation, and models in which the SFR or outflow efficiency
change rapidly compared to the $\sim 1\Gyr$ timescale of AGB
recycling could lead to significantly different $\dhr$ predictions.
For example, even with constant $\eta$ and $\xi_{\rm enh}=1$,
a 100 Myr starburst could eject most of the oxygen produced 
by its supernovae while allowing deuterium-depleted gas to return
from AGB envelopes after the burst has ended.
Consistent with this picture,
the models of \cite{dvorkin2016} that incorporate extreme early
star formation predict lower $\dhr$ than their smoother models.
We leave numerical investigation of bursty models with continuous
AGB recycling to future work.

Theoretical models of galaxy formation require feedback from
star formation to reproduce observed galaxy properties, and vigorously
star-forming galaxies at low and high redshift show abundant
evidence of outflowing gas.  However, there is little direct
observational evidence of winds in ``normal'' star-forming galaxies
like the Milky Way at $z=0$.  
\added{Many}
chemical evolution models suggest that
outflows with substantial mass-loading factors are required to produce
near-solar metallicity ISM abundances, as 
\added{
discussed in \S\ref{sec:bhcut} and}
illustrated in Figure~\ref{fig:bhcut}.
The high observed deuterium abundance in the local ISM
closes off two of the most
obvious loopholes in this argument: low IMF-averaged oxygen yields
and outflows that eject a large fraction of supernova metals
with little entrainment of the ambient ISM.

The robust link between oxygen and deuterium abundances should
also inform the lectures that we give to our introductory astronomy
astronomy students.  In the H$_2$O molecules that make up 2/3 
of our body mass, every oxygen atom was born in the nuclear
furnace of a stellar interior.  But the hydrogen atoms?
90\% of them come straight from the big bang.

\acknowledgements

\added{
I dedicate this paper to the memory of Gary Steigman, a pioneer in
the study of cosmic deuterium and its Galactic evolution, from
whom I learned much of what I know about the subject.}
I am also grateful to Bruce Draine and Marc Pinsonneault
for informative conversations about deuterium over the course of many years,
to Irina Dvorkin for a discussion of her results,
to Todd Thompson for comments on an early draft
of the manuscript, 
\added{
and to two anonymous referees whose questions and
suggestions led to significant improvements in the paper.
}
I also thank my GCE collaborators Brett Andrews, Jenna Freudenburg,
Jennifer Johnson, and Ralph Sch\"onrich for insights and valuable
background discussions.  This work was supported by NSF grant
AST-1211853.

\bibliographystyle{apj} 
\bibliography{deuterium}

%\begin{thebibliography}{999}
% contents of .bbl file
%\end{thebibliography}

\end{document}